# Optoelectronically Directed Self-Assembly of Active and Passive Particles into Programmable and Reconfigurable Colloidal Structures

*Donggang Cao[1,3], Sankha Shuvra Das[1,4], and Gilad Yossifon[1,2]\**

[1]School of Mechanical Engineering, Tel-Aviv University, Tel-Aviv, 69978, Israel

[2]Department of Biomedical Engineering, Tel-Aviv University, Tel-Aviv, 69978, Israel

[3]School of Engineering, University of the West of England, Bristol, BS16 1QY, United Kingdom

[4]Soft Condensed Matter, Raman Research Institute, Bangalore, 560080, India

*E-mail: gyossifon@tauex.tau.ac.il

Controlled assembly of active–passive colloidal mixtures offers a route to reconfigurable microscale machines, but their self-assembly pathways remain poorly understood. We study the directed assembly of metallo–dielectric Janus particles (JPs) and passive polystyrene (PS) beads using optoelectrically reconfigurable AC-field patterning, which allows precise control over particle composition and binding sequence. Through experiments, analytical modeling, and simulations, we show that dipolar interactions drive robust JP–JP and JP–PS dimer formation with frequency-dependent stability. At intermediate and high frequencies, JP–PS binding is strongly attractive, whereas at low frequencies it becomes effectively repulsive due to electrical double-layer screening and electrohydrodynamic flows at the metallic hemisphere. In multi-particle systems, PS beads act as cooperative hubs that hierarchically recruit JPs, yielding higher-order hybrid structures. We identify structural isomers—for example, 3JP + 1PS clusters can form chain-like or triangular configurations depending on assembly sequence. Simulations confirm both as equilibrium states, with the triangular isomer slightly more stable. Similar polymorphism appears in larger clusters (4JPs). Overall, we establish a framework for controlled active–passive colloidal assembly, showing how frequency-tunable interactions and structural polymorphism enable the design of reconfigurable colloidal machines for applications in microrobotics, targeted delivery, and adaptive materials.

## 1. Introduction

Self-propelling active particles have emerged as a versatile class of microrobotic platforms, with potential applications in targeted drug delivery [1], cell sorting [2], detoxification [3], immunosensing [4], remote surgery [5], and self-repairing systems [6]. Unlike passive colloids, which rely on externally imposed gradients such as dielectrophoresis (DEP) [7-8] or magnetophoresis [9-10], active particles autonomously convert locally harvested energy into directed motion. This propulsion arises from asymmetric designs that generate self-induced gradients, enabling mobility even under uniform conditions. Their intrinsic autonomy allows active particles to efficiently explore confined environments, cover larger spatial regions, and exhibit collective behaviours that drive higher-order self-organization [11]. These properties make them particularly attractive for biomedical microrobotics, where minimally invasive, dynamic, and precise manipulation is required [12-14].

Among the various active colloids investigated [15-20], metallo–dielectric Janus particles (JPs) are especially promising due to their straightforward fabrication, biocompatibility, and tuneable electrokinetic response. Under uniform alternating current (AC) electric fields, JPs propel via induced-charge electrophoresis (ICEP) or self-dielectrophoresis (sDEP) [21]. Field frequency serves as a powerful control parameter, regulating electrokinetic effects [22-23], directing locomotion, and tuning interactions with surrounding colloids and biological targets [24-25]. Thus, JPs can function simultaneously as autonomous micromotors and mobile microelectrodes, locally amplifying electric fields to enable advanced operations such as electroporation [26] and intracellular deformation [27]. These interactions can be understood as couplings between active JPs and passive objects (e.g., cells, nuclei, or colloids), where outcomes depend sensitively on field parameters, particle composition, and target properties.

Despite these advances, a fundamental limitation remains: all particles in the system respond collectively to the uniformly imposed electric field. In literature, most demonstrations are limited to controlling either single-particle micromotors [28-29] or colloidal swarms [30-32]. The former requires relatively dilute colloidal solutions to avoid uncontrolled interparticle interactions, which can lead to stochastic formation of colloidal assemblies and aggregates, e.g. chains of varying lengths.

To overcome these limitations, optoelectronic tweezers (OET) have emerged as a powerful tool for programmable and parallel manipulation of particles. OET utilize photoconductive substrates to convert projected light patterns into reconfigurable electric fields, enabling selective trapping, translation, and patterning of micro-objects with minimal optical power. Since its introduction by Chiou et al. [33-34], OET has demonstrated the ability to manipulate multiple particles simultaneously using optical images, providing a flexible, high-throughput platform for electrokinetic control. Subsequent developments—including patterned OET [35], OET-based microrobots [36], and OET-powered active colloids that exhibit frequency-dependent mobility reversal or guided self-assembly [37-38]—have shown that OET can direct both passive particle assemblies and active particle motion, enabling localized field patterning that overcomes the limitations of globally applied electric fields.

Although experiments have uncovered complex active–passive interactions, a systematic framework that links propulsion, interparticle interaction, and external fields to assembly outcomes remains limited. Here, we address the challenge by investigating the optoelectronically directed self-assembly of active–passive colloidal mixtures as a strategy for building dynamically reconfigurable hybrid structures, as shown in Figure 1. We couple the autonomous propulsion of JPs with PS bead interactions under AC fields, combining experiments, analytical models, and numerical simulations to uncover how field frequency, particle composition, and interaction history govern assembly. Our results reveal not only predictable binding motifs but also structural polymorphism in the form of order-dependent isomers, thereby establishing design rules for programmable and reconfigurable colloidal architectures. These findings have broad implications for the development of colloidal machines, enabling higher-throughput parallel assembly and the actuation of such structures with single cells for biomedical tasks such as targeted delivery and manipulation.

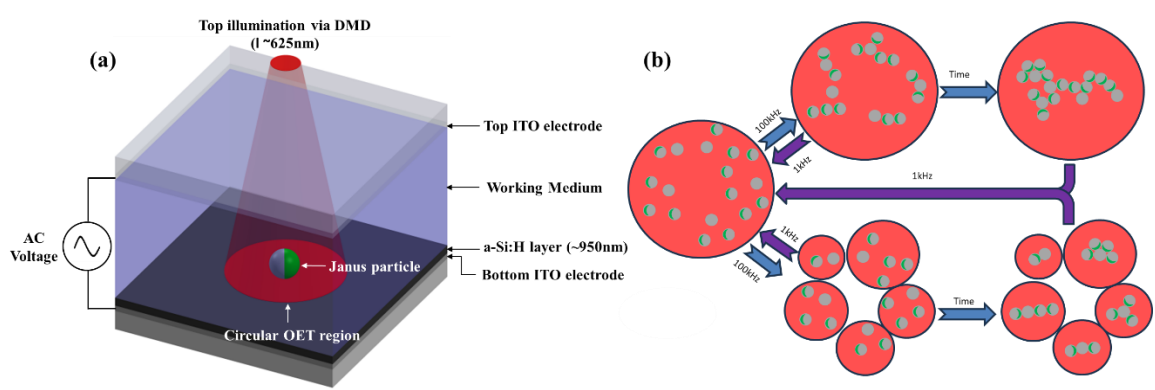

**Figure 1.** (a) Schematic of the microfluidic chamber used for optoelectronic (OE) manipulation of a Janus particle. The green and gray hemispheres represent the metallic and dielectric sides of the particle, respectively; (b) Schematic illustrating active–passive particle interactions under stochastic and optoelectronically guided conditions. Initially, Janus particles (JPs) and polystyrene (PS) beads are randomly dispersed within a large, uniformly illuminated optoelectronic tweezers (OET) region. Stochastic self-assembly: Under globally applied fields, small aggregates emerge randomly and gradually grow into larger, uncontrolled clusters. Optoelectronically programmed self-assembly: By projecting small, spatially defined OET patterns using a digital micromirror device (DMD), localized electric-field traps are generated, directing the formation of small, deterministic assemblies within each illuminated region. This approach enables precise, parallel, and reconfigurable construction of hybrid colloidal structures

## 2. Theoretical and Numerical Analysis

## 2.2 Analytical Modelling

Here we use the analytical expressions obtained in our previous publication [39] following also that of Yan *et al.* [40], wherein the dipolar electrostatic interaction potential between two particles, denoted A and B, can be expressed as

$$U_{AB} = 4\pi\varepsilon_0\varepsilon_s a^6 E_0^2 \frac{Re(K_A^* K_B)}{r_{AB}^3} \tag{1}$$

where $\varepsilon_0$ is the vacuum permittivity, $\varepsilon_s$ is the relative permittivity of the solution, $E_0$ is the amplitude of the applied electric field, $K_A = \frac{\varepsilon_A^* - \varepsilon_s^*}{\varepsilon_A^* + 2\varepsilon_s^*}$ and $K_B = \frac{\varepsilon_B^* - \varepsilon_s^*}{\varepsilon_B^* + 2\varepsilon_s^*}$ are the complex dipole coefficients of A and B, respectively ($\varepsilon^* = \varepsilon_0\varepsilon_r - \frac{\sigma}{\omega}i$ is the complex permittivity in which $\varepsilon_r$ indicates the relative permittivity, $\sigma$ represents the electric conductivity, $\omega$ is the angular frequency), and $r_{AB}$ is the distance between the induced dipoles.

Building on this formulation, we extend the method to systems involving multiple active Janus particles (JP) and passive polystyrene (PS) particles.

The total dipolar interaction energy for a JP–PS system in Figure 2 can be calculated by the equation below:

$$U = 4\pi\varepsilon_0\varepsilon_s a^6 E_0^2 \left[ \frac{Re(K_{A_d}^* K_{B_d})}{r_{A_d B_d}^3} + \frac{Re(K_{A_d}^* K_{B_m})}{r_{A_d B_m}^3} \right] \tag{2}$$

The subscripts *d* and *m* denote the dielectric and metallic hemispheres, respectively. The terms correspond to the interactions between the PS and JP's dielectric and metallic hemispheres, respectively. The dielectric properties of the used materials are listed in Table 1.

**Table 1 Dielectric properties of the Janus particle materials**

| Material | Relative permittivity | Electric conductivity (S/m) |
|---|---|---|
| Solution | 78.5 | 5×10$^{-4}$ |
| Polystyrene | 2.55 [11,41] | 1×10$^{-19}$ [11,42] |
| Gold | 1×10$^9$ [11] | 4×10$^7$ [11] |

With the dipole shift distance $d' = \frac{3}{8}a$ [40], the corresponding dipole–dipole separations at the orientation angle of $\theta_1$ are given by:

$$r_{A_d B_d} = \sqrt{(2a)^2 + \left(\frac{3a}{8}\right)^2 - 2(2a)\left(\frac{3a}{8}\right)\cos\theta_1} \tag{3}$$

$$r_{A_d B_m} = \sqrt{(2a)^2 + \left(\frac{3a}{8}\right)^2 - 2(2a)\left(\frac{3a}{8}\right)\cos(180° - \theta_1)} \tag{4}$$

Similarly, for a JP–JP pair in Figure 2, the total dipolar interaction energy can be expressed as the sum of four pairwise dipole–dipole interactions:

$$U = 4\pi\varepsilon_0\varepsilon_s a^6 E_0^2 \left[\frac{Re(K_{A_d}^* K_{B_d})}{r_{A_d B_d}^3} + \frac{Re(K_{A_d}^* K_{B_m})}{r_{A_d B_m}^3} + \frac{Re(K_{A_m}^* K_{B_d})}{r_{A_m B_d}^3} + \frac{Re(K_{A_m}^* K_{B_m})}{r_{A_m B_m}^3}\right] \quad (5)$$

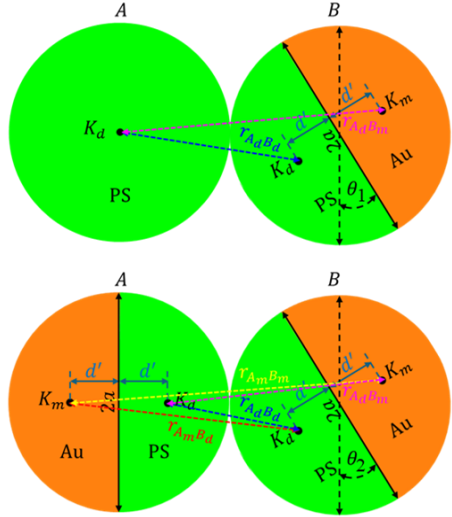

**Figure 2.** Schematic representation of dipolar interactions (top: JP+PS; bottom: JP+JP).

The terms correspond to the interactions between dielectric-dielectric, dielectric-metallic, metallic-dielectric, and metallic-metallic hemispheres of the two JPs. The effective dipole–dipole distances entering Eq. (5) are determined by simple geometric relations, summarized as follows:

$$r_{A_d B_d} = \sqrt{\left(a + \frac{5a}{8}\right)^2 + \left(\frac{3a}{8}\right)^2 - 2\left(a + \frac{5a}{8}\right)\left(\frac{3a}{8}\right)\cos\theta_2} \quad (6)$$

$$r_{A_d B_m} = \sqrt{\left(a + \frac{5a}{8}\right)^2 + \left(\frac{3a}{8}\right)^2 - 2\left(a + \frac{5a}{8}\right)\left(\frac{3a}{8}\right)cos(180° - \theta_2)} \quad (7)$$

$$r_{A_m B_d} = \sqrt{\left(2a + \frac{3a}{8}\right)^2 + \left(\frac{3a}{8}\right)^2 - 2\left(2a + \frac{3a}{8}\right)\left(\frac{3a}{8}\right)cos(\theta_2)} \quad (8)$$

$$r_{A_m B_m} = \sqrt{\left(2a + \frac{3a}{8}\right)^2 + \left(\frac{3a}{8}\right)^2 - 2\left(2a + \frac{3a}{8}\right)\left(\frac{3a}{8}\right)cos(180° - \theta_2)} \quad (9)$$

In the formulation, $\theta_2$ represents the angular orientation of the interacting dipoles.

The expressions highlight how the dipolar interaction depends sensitively on both the material properties (through the dipole coefficients $K$) and the geometric configuration of the particles (through $r$ and $\theta_1, \theta_2$).

## 2.3 Numerical Simulations

The numerical simulation of the electrostatic problem was carried out in COMSOL 6.3. A three-dimensional geometry was constructed to represent the experimental setup, consisting of a block channel (120 µm height, 170 µm width, and 80 µm length) with 27-µm-diameter spheres positioned 60 µm above the substrate.

Under the weak-field assumption, where the applied voltages are small compared with the thermal potential, surface conduction and convection within the electric double layer (EDL) can be neglected. The EDL may therefore be modeled as a linear capacitor. Since the EDL thickness ($\lambda$) is much smaller than the particle radius ($a$), i.e., $\lambda/a \ll 1$, the Laplace equation for the electric potential, $\nabla^2 \phi = 0$, can be solved for the electric potential, $\phi$, subject to the following boundary condition at the metallic side of the Janus particle [7]:

$$\sigma \frac{\partial \phi}{\partial n} = i\omega C_{DL}(\phi - V_{floating}) \qquad (10)$$

This condition describes the oscillatory Ohmic charging of the induced EDL, where $V_{floating}$ is the floating potential of the metallic hemisphere, $n$ is the outward normal to the JP surface, and $C_{DL}$ is the capacitance per unit area of the EDL, estimated using the Debye–Hückel theory as $C_{DL} = \varepsilon/\lambda$. Same values of permittivity and conductivity (Table 1) are used in both numerical and analytical studies. However, in the simulation a floating boundary condition [38] was imposed on the metallic hemisphere to enforce zero net charge. The dielectric hemisphere of the JP was treated with an insulation boundary condition. A voltage of 20 V was applied at the lower substrate ($z = 0$), the upper wall was grounded, and the channel sidewalls were assigned insulating boundary conditions. In the numerical simulations we have varied both the gap between the interacting particles and the orientation angle, $\theta_1(\theta_2)$ (defined in Figure 2), of the metallo-dielectric interface of the examined JP when calculating the total stored electric energy in the system [11,43].

## 3. Results and Discussion

### 3.1. Nature of particles assembly

The assembly behaviour of active–passive and active–active particle pairs under an applied AC electric field was systematically analyzed to elucidate the fundamental mechanisms governing their interactions. Figure 3 provides a comprehensive comparison of experimental observations, analytical predictions, and numerical simulations.

Figure 3(A) presents representative experimental images of the two assembly types (see Movie 1 in SI). In the active–passive system, a metallo-dielectric Janus particle (JP) interacts with a polystyrene (PS) particle, forming a stable dimer configuration in which the gold hemisphere of the JP connects to the PS particle. In the active–active system, two JPs self-assemble into chain-like structures, with the gold hemisphere of one JP facing the polystyrene hemisphere of the other. The corresponding schematics illustrate the orientation of the metallic and dielectric hemispheres relative to the applied field (**E**).

Using an analytical charged-dipole model (see Materials and Methods section), Figure 3(B) shows the electric energy at fixed dipole charge as a function of particle orientation angle, $\theta_1/\theta_2$ (see Fig. 2). For both active–passive and active–active systems at zero gap distance (i.e., contacting particles), the electric energy exhibits a pronounced minimum at $\theta_1(\theta_2) = 180°$, corresponding to an energetically favorable head-to-tail alignment. This indicates that electrostatic interactions intrinsically drive the formation of linear assemblies, with the energy minima reflecting the preferred relative orientations in agreement with the experimental observations in Figure 3(A).

Figure 3(C) further illustrates the dependence of electric energy on interparticle gap distance for particles maintained in head-to-tail alignment. As the normalized gap distance ($l/a$, where $a$ is the particle radius) increases, the electric energy rises monotonically, indicating decreasing stability. This trend is observed for both active–passive and active–active pairs. Notably, at short distances ($l/a < 1$), the energy changes sharply, revealing increasing stability at close interparticle proximity. At larger separations ($l/a > 1.5$), the energy approaches an asymptotic value, indicating that electrostatic coupling weakens with distance.

In the experimental setup, the electric field is generated between the top and bottom substrate electrodes. The bottom electrode is optically patterned, consisting of a photoconductive layer deposited on an ITO-coated glass substrate, and both electrodes are maintained at fixed potentials via an external power source.

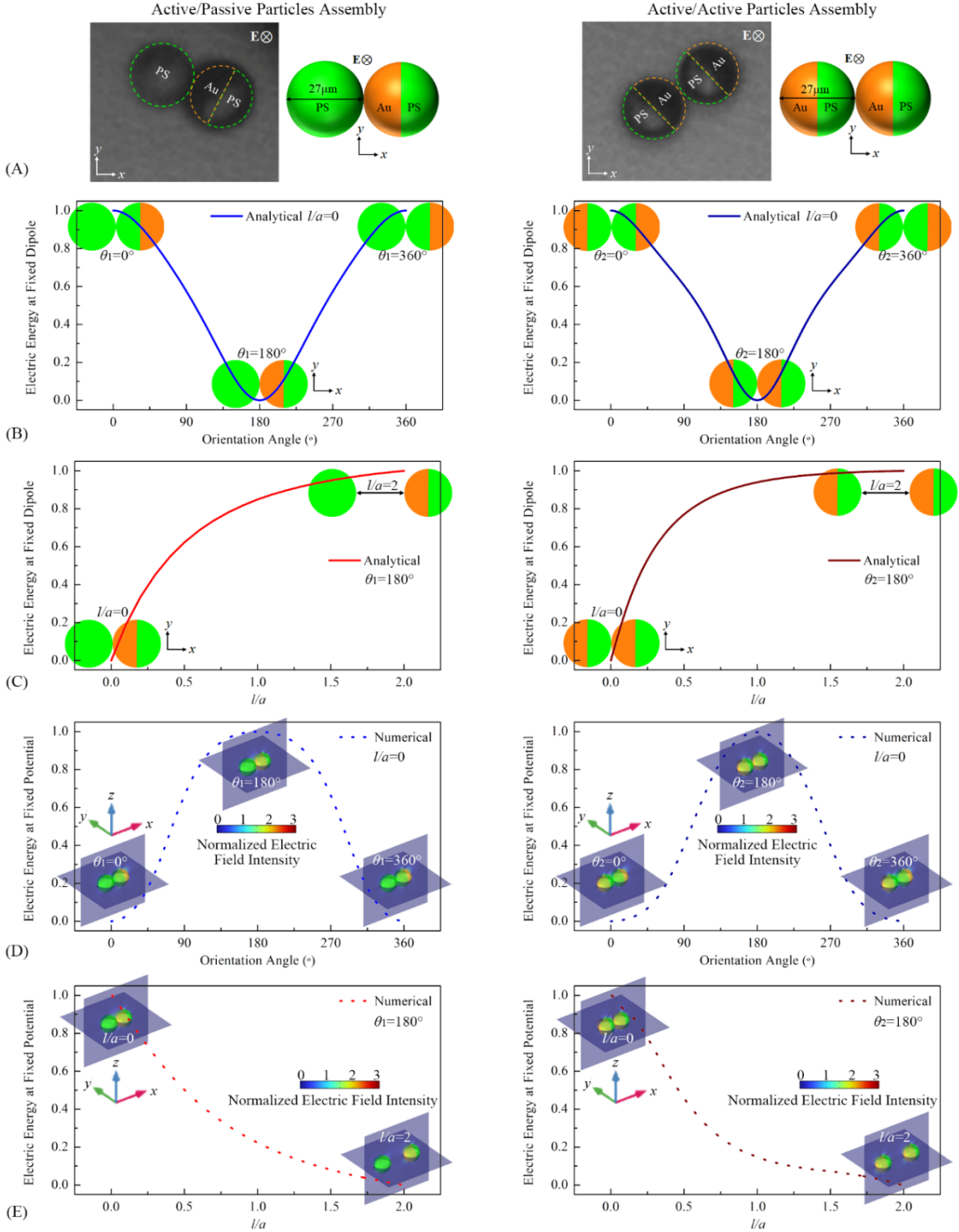

**Figure 3.** Assembly of active–passive and active–active particle pairs under an AC electric field ($f$=10 kHz, $V_{pp}$=20 V): (A) Experimental images and schematics showing a JP–PS (left) and JP–JP dimers (right). (B–C) Analytical energy landscapes at fixed dipole charge as a function of (B) orientation angle, $\theta_1(\theta_2)$, and (C) interparticle gap distance, $l/a$ (normalized by the particle radius $a$) revealing preferred head-to-tail alignment. (D–E) Numerical energy landscapes at fixed potential as a function of (D) orientation angle and (E) gap distance, confirming the stabilization of head-to-tail assemblies.

As the system's dielectric polarization varies with colloidal configurational changes, charge flows to or from the electrodes to maintain the fixed potential, with the required energy supplied by the external source. Consequently, the equilibrium configuration corresponds to the maximum stored electric energy in reality [44]. Although numerical simulations of electric energy at fixed potential exhibit the opposite trend from fixed charge system solved analytically, they reach maximum values at $\theta_1(\theta_2) = 180°$ (Figure 3(D)) and zero gap (Figure 3(E)), corroborating the analytical predictions and providing insight into the three-dimensional distribution of normalized electric field intensity.

While numerical (at fixed potentials) and analytical (at fixed charges) results display opposite trends in absolute energy values, both approaches consistently demonstrate that the electric energy governs the assembly of active–passive and active–active pairs, with equilibrium configurations corresponding to extremal points. Collectively, these results indicate that the assembly of colloidal particles under AC fields is primarily controlled by orientation-dependent dipolar interactions and distance-dependent electrostatic coupling. Preferential head-to-tail alignment, underlies the formation of both active–passive and active–active structures. In the following sections, we use numerical simulations that explicitly incorporate battery effects, providing a faithful representation of the experimental setup and enabling analysis of more complex assembly scenarios.

### 3.2. Frequency impact on the active/passive particles interaction

The electrostatic interaction between particles and in particular between a metallo-dielectric Janus particle (JP) and a passive polystyrene (PS) bead is strongly dependent on the excitation frequency. Figure 4 consolidates the experimental images, numerical electric field, and computed electric energy landscapes at fixed potential for the 1JP + 1PS system across the frequency range studied. Together these data show that frequency not only controls the magnitude of the induced polarization on the JP, but can also change the sign of the JP–PS interaction (repulsive ⇌ attractive) and thereby determine the resulting interparticle gap and orientation of the colloidal pair.

Figure 4(A) presents representative experimental snapshots at 10 kHz and 100 kHz. In both images the JP and PS form a stable dimer with the gold hemisphere of the JP in contact with the PS particle. Frequencies below 1 kHz were not examined to avoid possible water electrolysis and degradation of the electrodes and microfluidic chamber [7].

Figure 4(B) displays numerically calculated normalized electric field for several frequencies, with the induced EDL accounted for in the boundary conditions (Eq. (10)). Two regimes are evident. At lower frequencies (e.g., 0.1 kHz) relative to the induced charge relaxation frequency ($f_{RC}$=1/2$\pi\tau$=447Hz, where $\tau=\lambda a/D$ is the induced charge relaxation time; $a$=13.5μm is the radius of JP; $\lambda = \sqrt{\varepsilon D/\sigma}$ is the Debye length; $\sigma$=0.5mS/m is the solution conductivity; $\varepsilon = 78.5\varepsilon_0$ is the solution permittivity and $\varepsilon_0$ =8.854187817×10$^{-12}$F/m is the vacuum permittivity; $D$=2×10$^{-9}$m$^2$/s is the diffusion coefficient of the ionic species [38]), the EDL has sufficient time to charge and screen the metallic hemisphere; under these conditions the JP's electrostatic response, externally to its thin induced EDL, resembles that of the dielectric PS hemisphere and the contrast between the Au and PS sides is small. Conversely, at frequencies above the RC frequency (i.e. 1 kHz and above), EDL charging cannot keep pace with the field oscillations, leading to progressively weaker screening, and the Au hemisphere increasingly behaves as a good conductor.

The numerical electric field distribution show that, as frequency increases, the local electric field concentrates progressively at the Au–PS interface.

Figures 4(C) and 4(D) quantify how these shifts in electric field distribution map onto the interaction energy. It is worth noting again that under the constant imposed potential the configurations corresponding to larger electric energy are more stable. Figure 4(C) shows the electric energy at fixed potential as a function of orientation angle (for contact $l/a$=0). For all frequencies the orientation (Au facing PS, head-to-tail) corresponds to an energetic extremum and therefore to the preferred bound orientation. The amplitude of this energy extremum increases with frequency once the system is above the RC frequency, consistent with the change of the electric field distribution shown in Fig. 4(B).

Crucially, Figure 4(D) - electric energy at fixed potential for $\theta_1 = 180°$ as a function of normalized gap $l/a$ - reveals opposite interparticle gap-dependences in the low- and high-frequency regimes. At low frequencies (0.1 kHz and near $f_{RC}$), the JP behaves electrically similar to a PS particle and the interaction is effectively repulsive (i.e. the computed electric energy increases with increasing gap distance). Because higher energy denotes greater stability in our convention, this trend indicates the stable configuration in the low-frequency regime is a separated state (the PS particle tends to stay away from the JP).

By contrast, at intermediate and high frequencies (≥1 kHz, including the experimentally probed 10 kHz and 100 kHz), EDL screening weakens and the Au hemisphere behaves as a conductor; the interaction becomes attractive, and the computed energy decreases with increasing gap. In this regime the energy is maximal at contact ($l/a\rightarrow$0), indicating that the most stable configuration is the bound JP–PS dimer. The numerical results thus explain the experimental observation in Fig. 4(A): at both 10 kHz and 100 kHz, the JP preferentially binds to the PS particle through its Au hemisphere. At 1 kHz, the numerical simulations indicate a weak attractive interaction, whereas experimentally the particles show no attraction (See Movie 5 in SI) due to electrohydrodynamic effects—absent in the simulations, which considered only electrostatic forces.

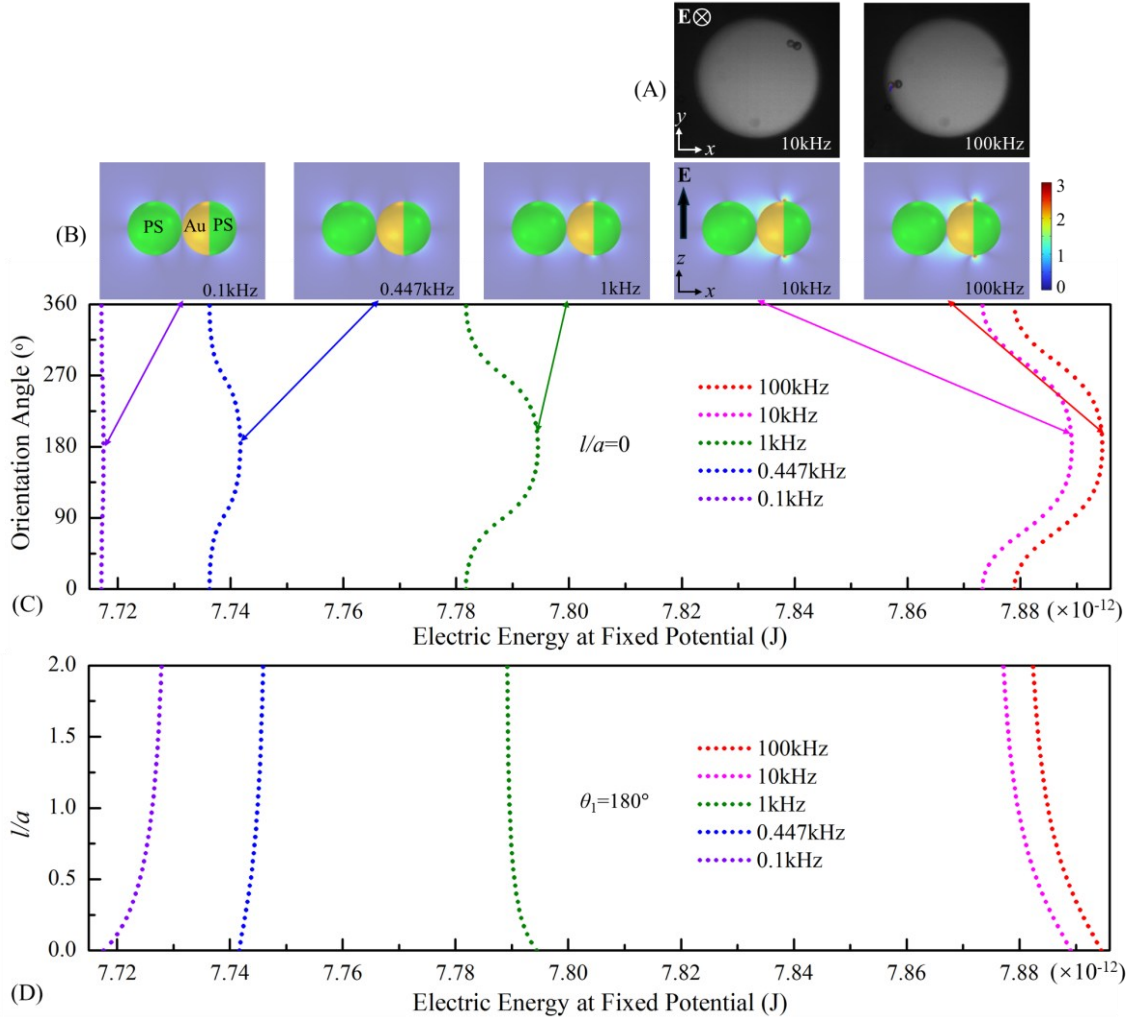

**Figure 4.** Frequency-dependent numerically computed electrostatic interactions in the 1 Janus particle (JP) + 1 polystyrene (PS) particle system ($V_{pp}$=20 V): (A) Experimental snapshots at 10 kHz and 100 kHz showing stable JP–PS dimers. (B) Numerical electric field maps (normalized by $E_0$=166.7kV/m) illustrating low-frequency induced electric-double-layer (EDL) screening and high-frequency electric field local intensification at the Au–PS interface. (C–D) Electric energy at fixed potential as a function of (C) orientation angle, $\theta_1$, and (D) interparticle gap distance, $l/a$ (normalized by the particle radius $a$=13.5μm), demonstrating the transition from repulsive interactions at low frequency to attractive binding at intermediate and high frequencies.

### 3.3. Interactions between multiple active particles and a single passive particle

The collective dynamics of multiple JPs interacting with a passive polystyrene (PS) bead introduce additional complexity beyond the pairwise assemblies discussed earlier. Figure 5 consolidates experimental time-sequence images with numerical calculations of electric energy at fixed potential, capturing both the temporal evolution of cluster formation and the energetic mechanisms that stabilize these colloidal assemblies.

Figure 5A presents experimental time-evolving snapshots illustrating the dynamics of the 2 JP + 1 PS system at 10 kHz (see Movie 2 in the SI). In Sequence I, the polystyrene (PS) particle is initially far from the two Janus particles (JPs), which first assemble through contact between the gold hemisphere of one JP and the polystyrene hemisphere of the other, consistent with the previously observed active–active assembly mode (Fig. 3). In Sequence II, the PS particle and the JP dimer subsequently approach each other. In Sequence III, the PS particle migrates toward the JP dimer and transiently interacts with both JPs. In Sequences IV and V, the PS particle stabilizes at the center, bridging the two JPs, with both Au hemispheres in direct contact with the PS particle.

The corresponding numerical results in Figure 5(B) quantify the increase in electric energy at fixed potential as the 2JPs + 1PS system evolves through these configurations. The gradual increase in energy from states I to V reflects the progressive stabilization of the assembly as more favorable colloidal configuration, where the PS sits symmetrically between the two JPs with Au hemispheres in direct contact, corresponding to the highest energy state and therefore to the most stable equilibrium configuration.

Figure 5(C) extends this analysis to higher-order assemblies by computing the electric energy as a function of orientation angle of each subsequently introduced JP in progressively larger JP–PS clusters. In the first subfigure, the PS is fixed and a single JP rotates, equivalent to the active/passive case in Fig. 3. The energy curve shows that the system is most stable when the Au hemisphere faces and connects to the PS (orientation angle $\theta_1 = 180°$). In the second subfigure, the equilibrium configuration of a JP–PS dimer (Au of JP bound to PS) is treated as a unit, and a second JP is added to the left and allowed to rotate. The results again show that stability is maximized when the Au hemisphere of the rotating JP binds to the PS. Similarly, in the third and fourth subfigures, the equilibrium configurations of 2JPs + 1PS and 3JPs + 1PS are treated as units, with additional JPs added sequentially and rotated. In each case, the energy maxima occurs when the Au hemisphere of the newly added JP connects to the polystyrene hemisphere of its neighbor JP. These results reveal a cooperative and hierarchical assembly process. The energy landscapes confirm that each newly added JP follows the same binding rule, yielding predictable, dynamically reconfigurable hybrid colloidal structures centered on the PS. The trajectories of assembled structures using different particles can be seen in Movie 3 in SI.

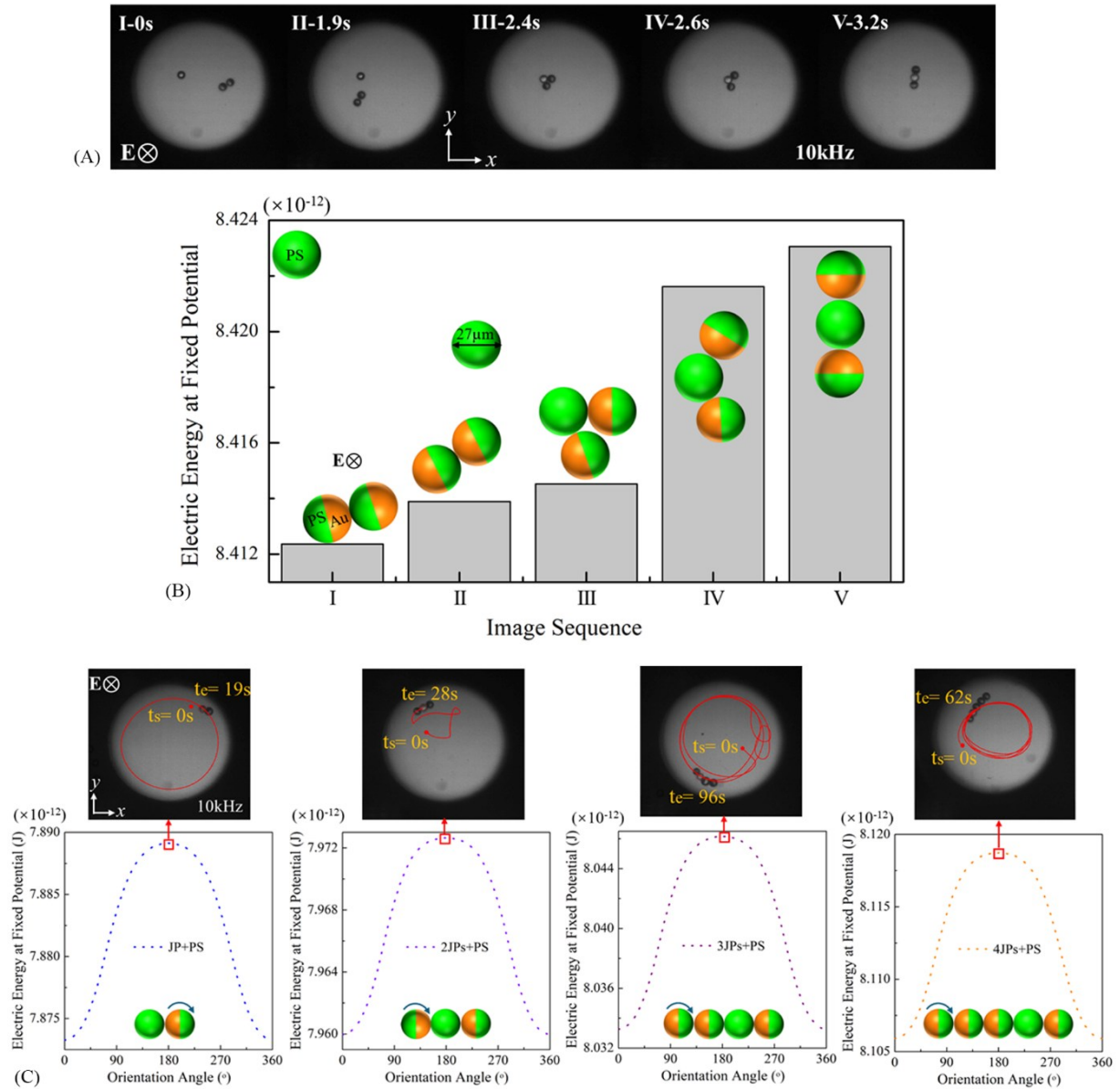

**Figure 5.** Cooperative assembly of multiple active Janus particles (JPs) with a single polystyrene (PS) passive particle ($f$=10 kHz, $V_{pp}$=20 V): (A) Experimental time sequence of a 2JPs + 1PS system, showing stepwise recruitment of the PS bead into the JP dimer and stabilization of a symmetric three-body structure. (B) Numerical evolution of the electric energy at fixed potential approximately correlates the experimentally observed, time evolving structural configurations - each numerically modeled at its instantaneous state - with increasing stability. (C) Simulated energy landscapes for progressively larger JP–PS clusters, with varying orientation angle of the consequentially added JP, confirming hierarchical recruitment of JPs via Au–PS contacts and the role of PS as a cooperative hub.

### 3.4. Structural isomers of active/passive particles interactions

The interactions of multiple active JPs with a single passive PS bead may yield varying structural isomers—distinct stable configurations assembled from the same components. Figure 6 illustrates this in the 3JPs + 1PS system, where both a chain-like structure and a triangular structure emerge (see Movie 4 in SI). Crucially, the formation of these isomers is history-dependent, as the sequence in which individual JPs bind to the PS determines the final architecture.

Figure 6(A) presents experimental time-resolved images of two representative assembly pathways. In pathway I, the process begins with a 1JP + 1PS dimer (I-1), consistent with earlier observations of active–passive binding (Sections 3.1), where the PS preferentially attaches to the Au hemisphere of a JP. Two additional JPs remain separated at this stage. In I-2, the free JPs approach the dimer, and in I-3, one JP attaches to the existing JP rather than to the PS, producing a 2JPs + 1PS complex with one JP still detached. Finally, in I-5, the last JP binds to the PS, completing a linear 3JPs + 1PS chain. This pathway demonstrates how sequential JP-to-JP and JP-to-PS interactions bias the assembly toward a chain isomer.

Pathway II, by contrast, yields the triangular isomer through a different order of events. Similar to pathway I, the PS first binds to a JP at the Au hemisphere (II-1). However, in II-2, the second JP binds directly to the PS, positioning the passive particle centrally—analogous to the cooperative hub behavior described in Figure 5(A-B), where PS recruited two JPs symmetrically. As the last JP migrates inward (II-5), it also binds to the PS, forming a triangular 3JPs + 1PS configuration in which the passive particle anchors all three JPs.

Figure 6(B) presents numerical results of the electric energy at fixed potential over time at excitation frequencies of 10 kHz and 100 kHz, showing that the overall energy is higher at higher frequency. This trend is consistent with the frequency-dependent stability discussed in Section 3.2, where stronger polarization of the partially screened Au hemisphere enhances binding. Both the chain and triangular configurations correspond to stable energy maxima, confirming both as possible equilibrium isomers. Notably, the triangular configuration has slightly higher energy than the linear chain.

In addition to their distinct energy landscapes, the two isomers display clearly different dynamical behaviors (Fig. 6C). At 10 kHz, the chain isomer traces a smooth circular trajectory within the illuminated region, staying well inside the ring boundary. At 100 kHz it instead remains near the ring edge with only slight movement, indicating a strong reduction in propulsion at higher frequency. The triangular isomer shows the same qualitative trend: at 10 kHz it moves within the circular illumination but with markedly weaker propulsion, consistent with its nearly symmetric geometry, and at 100 kHz it becomes effectively immobilized near the ring edge with negligible motion.

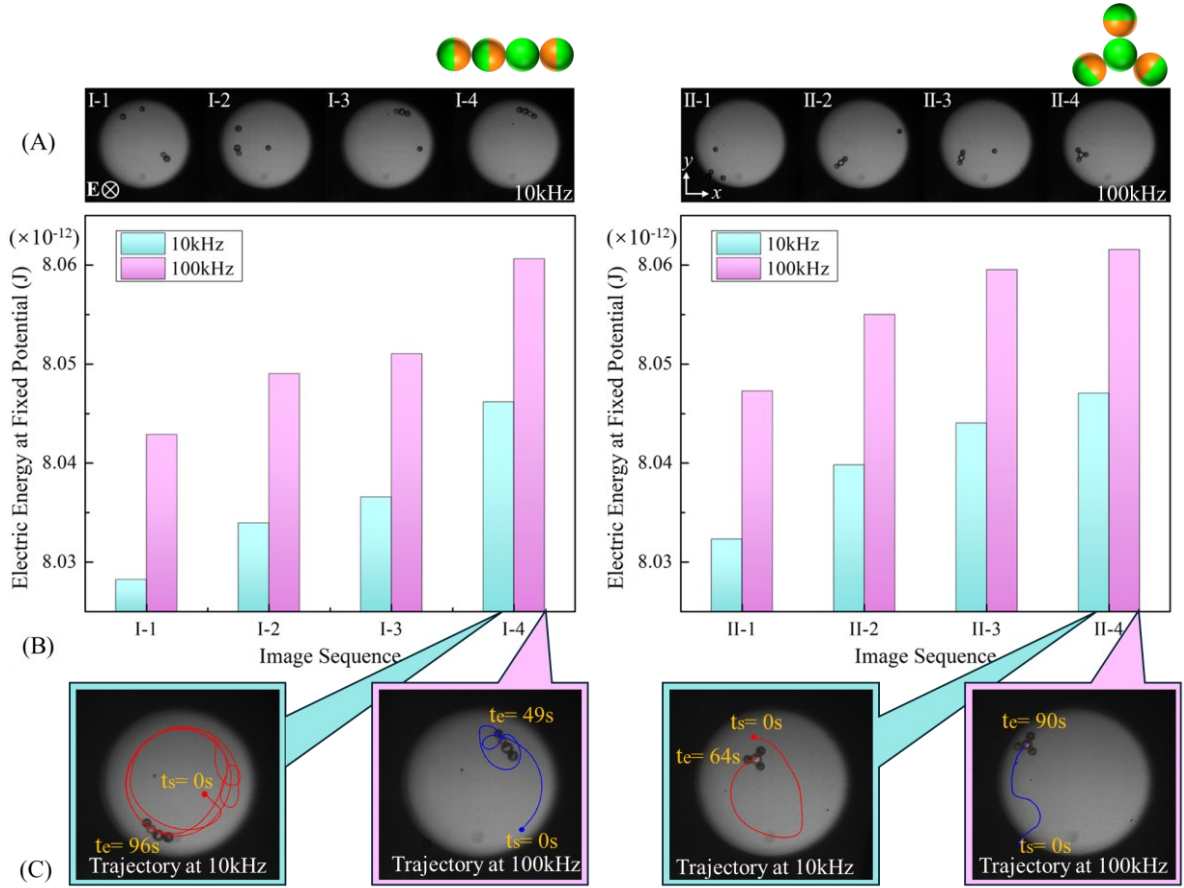

**Figure 6.** Structural isomers in the 3 Janus particles (JPs) + 1 polystyrene (PS) system. (A) Experimental assembly pathways yielding either a chain-like or triangular colloidal configurations, depending on binding order. (B) Numerical electric energy at fixed potential over time at 10 kHz and 100 kHz, showing frequency-enhanced stability and confirming both chain and triangular isomers as equilibrium states, with the triangular geometry slightly energetically favored. (C) Distinct trajectories of the two structural isomers at 10 kHz and 100 kHz. At 10 kHz, both isomers exhibit noticeable circular motion within the illuminated region, whereas at 100 kHz they remain near the edge of the optical circular region with reduced displacement.

Taken together, these findings reveal that structural polymorphism in active–passive assemblies is dependent on assembly order and interaction history. The PS particle consistently serves as a cooperative hub that anchors and recruits successive JPs, but the order of binding dictates whether the system stabilizes in a chain or triangular geometry. Notably, similar structural isomers have also been experimentally observed for larger assemblies involving 4JPs

+ 1PS, as shown in Figure S1 in the Supporting Information. This history-dependent structural diversity mirrors molecular isomerism in chemistry and provides a pathway to program reconfigurable colloidal architectures with tunable and switchable colloidal structures and function.

## 4. Conclusion

In this study, we have elucidated the mechanisms underlying the optoelectronically directed self-assembly of active–passive colloidal systems driven by AC electric fields. By integrating experimental observations with analytical and numerical models, we established how frequency-dependent dielectric polarization governs the attractive or repulsive character of JP–PS interactions and how passive beads act as cooperative hubs that recruit successive active particles. This hierarchical assembly produces not only predictable hybrid structures but also structural polymorphism: identical particle sets can stabilize into distinct stable isomeric states depending on their assembly history. As an example, the 3JPs + 1PS system exhibits both chain-like and triangular configurations, with numerical results confirming their equilibrium stability and experiments validating their time-evolving assembly dynamics. These results highlight that active–passive mixtures can encode both tunability and memory, where the assembly pathway determines the final configuration. Extending to larger JP–PS clusters, the observation of structural isomers further underscores the potential for programmable polymorphism. These insights provide a generalizable design principle for constructing dynamically reconfigurable colloidal architectures, and open opportunities for dynamically reconfigurable microrobotic platforms that can switch between multiple structural states, adapt to changing environments, and execute complex functions at the single-cell level.

## 5. Experimental Section

*Fabrication of Janus Particles and Solution Preparation:*

Polystyrene microspheres (27 μm in diameter, Sigma-Aldrich) were dispersed in isopropanol (IPA) and drop-cast onto piranha-cleaned glass substrates. Following solvent evaporation, the particles formed a close-packed monolayer. A thin metal trilayer was then sequentially deposited—15 nm of chromium, 50 nm of nickel, and 15 nm of gold—following the protocol of Wu *et al.* [26-27]. The coated particles were detached from the substrate by sonication in deionized water (DIW) containing 2% (v/v) Tween-20 (Sigma-Aldrich). The resulting Janus particles (JPs) were rinsed three times with the working medium to minimize substrate adhesion.

The working medium consisted of 10 μM KCl solution (prepared by serial dilution from a 1 M stock) with 0.1% (v/v) Tween-20, corresponding to a conductivity of ~5 μS/cm.

*Optoelectronic Microfluidic Device Fabrication and System Setup:*

The optoelectronic (OE) microfluidic chamber was constructed by sandwiching a ~120 μm-thick spacer (Grace-Bio) between two indium tin oxide (ITO)-coated glass substrates (200 nm ITO layer, Sigma-Aldrich) (Figure 1a). The bottom ITO substrate was coated with 900-1000 nm of hydrogenated amorphous silicon (a-Si:H) by plasma-enhanced chemical vapor deposition (PECVD, Oxford Plasma Instruments, Plasma Lab 100), using the following deposition recipe: $SiH_4$ (20 sccm) and $H_2$ (60 sccm) at 200 °C, 1000 mTorr, and 30 W RF power. The electrical conductivity of the a-Si:H film was characterized by measuring current under 5 V applied bias under dark and illuminated conditions, giving dark and photo-conductivities of $\sim 10^{-10}$ and $\sim 10^{-7}$ $Sm^{-1}$, respectively. Electrical contacts were made using copper tapes (3M9887-ND, Digi-Key) attached to the ITO surfaces. AC voltages were applied between the two electrodes using a function generator (Agilent 33250A) and monitored with an oscilloscope (Tektronix TPS-2024). The custom-built OET setup integrates a DMD-based illuminator (Texas Instruments, DLP2000 EVM) with an upright microscope (Olympus, model: U-ST), enabling projection of spatial light patterns onto the OE substrate (Figure 1a. A convex lens (Thorlabs, LB1757-A) positioned between the DMD and the microscope port focuses the LED output ($\lambda \approx 625$ nm, power $\approx 0.23$ W) onto a dichroic mirror, which directs the patterned illumination through a 10× objective lens onto the substrate mounted on a manual XY stage (Thorlabs, PT1). The DMD was programmed to project stationary, pre-defined image patterns. For imaging, a CMOS camera (Thorlabs, DCC1645C) was attached to the microscope's top port, with a short-pass filter (Thorlabs, FES0550; cut-off $\lambda \approx 550$ nm) placed in front to prevent pixel saturation from reflecting red light. The optical power density at the substrate plane, corresponding to a circular illumination spot projected via the 10× objective, was measured as $\approx 0.051$ W $cm^{-2}$ using an Ophir Photonics VEGA power meter.

**Supporting Information**

Supporting Information is available from the Wiley Online Library or from the author.

**Acknowledgments**

G.Y. acknowledges support from the Israel Science Foundation (ISF) (1429/24). We thank the Center for Nanoscience and Nanotechnology, Hebrew University of Jerusalem for assisting us in fabricating the photoconductive substrates.

**Supporting Information**

**Optoelectronically Directed Self-Assembly of Active and Passive Particles into Programmable and Reconfigurable Colloidal Structures**

*Donggang Cao[1,3], Sankha Shuvra Das[1,4], and Gilad Yossifon[1,2]\**

[1]School of Mechanical Engineering, Tel-Aviv University, Tel-Aviv, 69978, Israel

[2]Department of Biomedical Engineering, Tel-Aviv University, Tel-Aviv, 69978, Israel

[3]School of Engineering, University of the West of England, Bristol, BS16 1QY, United Kingdom

[4]Soft Condensed Matter, Raman Research Institute, Bangalore, 560080, India

**Figures:**

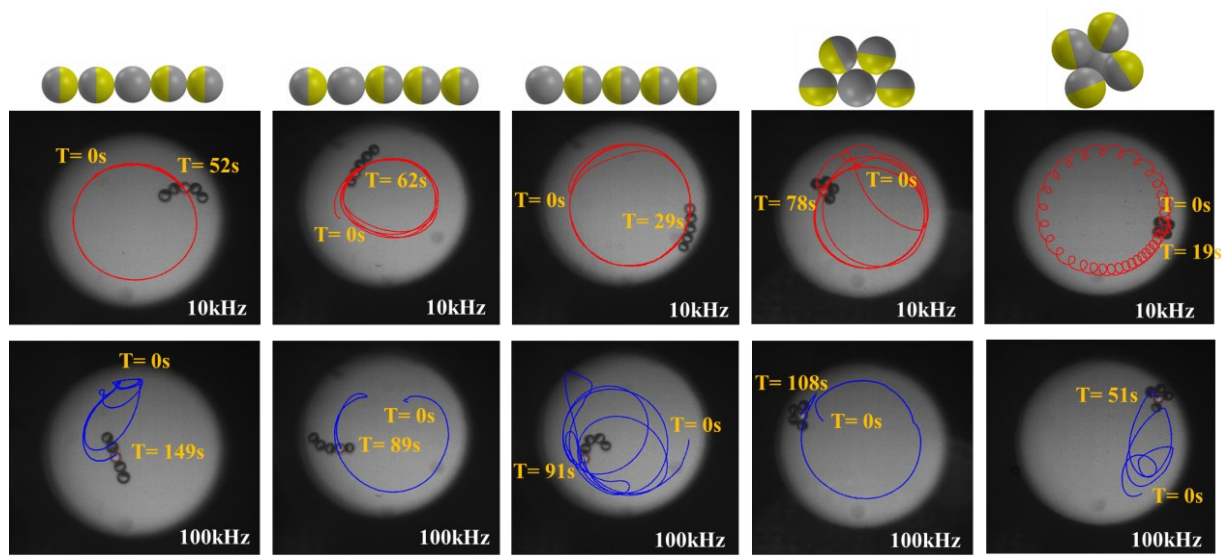

**Figure S1.** Structural isomers, which is highly dependent on the order history, in the 4 Janus particles (JPs) + 1 polystyrene (PS) system with distinct trajectories at 10 kHz and 100 kHz.

**Movies:**

Movie 1: Dynamic assembly of 1JP+1PS and 1JP+1JP (to support Figure 3).

Movie 2: Dynamic assembly of 2JP+1PS (to support Figure 5A).

Movie 3: Trajectories of 1JP+1PS, 2JPs+1PS, 3JPs+1PS, 4JPs+1PS at 10 kHz (to support Figure 5C).

Movie 4: Dynamic assembly and trajectories of different structural isomers of 3JPs+1PS (to support Figure 6).

Movie 5: Assembly (20$V_{pp}$, 100 kHz) and disassembly (20$V_{pp}$, 1 kHz) of JP and PS and particles.

21